\documentclass[astrosymb,twocolumn]{aastex701}
\usepackage[T1]{fontenc}
\usepackage{graphicx}
\usepackage{amsmath}
\usepackage{amssymb}
\usepackage{bm}
\usepackage{hyperref}
\hypersetup{
  colorlinks   = true,
  citecolor   = blue
}
\usepackage{natbib}
\usepackage{color}
 
\newcommand{\beq}{\begin{equation}}
\newcommand{\eeq}{\end{equation}}

\newcommand{\benum}{\begin{enumerate}}
\newcommand{\eenum}{\end{enumerate}}

\def\apjs{ApJS}
\def\aap{A \& Ap}

\def\apjl{ApJL}
\def\aap{A\&A}

\def\bar{\overline}
\newcommand{\abra}[1]{\left\langle{#1}\right\rangle}

\newcommand{\rin}{r_0}
\newcommand{\rout}{r_\text{out}}
\newcommand{\cs}{c_\text{s}}
\newcommand{\cso}{c_{\text{s}0}}

\newcommand{\Msun}{M_\odot}

\newcommand{\rg}{r_\text{g}}

\newcommand{\taud}{\tau_\text{d}}

\newcommand{\omegaD}{\omega_\text{LSD}}
\newcommand{\kD}{k_\text{LSD}}

\newcommand{\Mdot}{\dot M}

\newcommand{\MBH}{M_\text{BH}}

\newcommand{\CO}{C_\Omega}
\newcommand{\Cl}{C_l}
\newcommand{\Cbeta}{C_\beta}

\newcommand{\etaEdd}{\eta_\text{Edd}}
\newcommand{\LEdd}{L_\text{Edd}}
\newcommand{\xiin}{r_0}
\newcommand{\xiout}{\xi_\text{out}}

\newcommand{\Clyu}{C_\text{L97}}

\newcommand{\alphaLSD}{\tilde\alpha_\text{LSD}}

\begin{document}

\title{Understanding the UV/Optical Variability of AGNs through Quasi-Periodic Large-scale Magnetic Dynamos}

\author[0000-0002-2991-5306]{Hongzhe Zhou}
\email[show]{hzzhou@sspu.edu.cn}
\affiliation{Department of Physics, School of Mathematics, Physics and Statistics, Shanghai Polytechnic University, 2360 Jin Hai Road, Shanghai, 201209, China}
\affiliation{Tsung-Dao Lee Institute, Shanghai Jiao Tong University, Shanghai, 201210, China}

\author{Dong Lai}
\email[show]{dong.lai@sjtu.edu.cn}
\affiliation{Tsung-Dao Lee Institute, Shanghai Jiao Tong University, Shanghai, 201210, China}
\affiliation{Department of Astronomy and Cornell Center for Astrophysics and Planetary Science, Cornell University, Ithaca, NY 14853, USA}

\begin{abstract}

The physical origin of the the recently identified slow-moving temperature fluctuations in accretion disks around super-massive black holes (SMBHs) cannot be accounted for by reverberation models.
In this work, we propose that large-scale dynamos (LSDs) operating in accretion disks could generate quasi-periodic perturbations in the turbulence viscosity, thereby producing outward-going temperature fluctuations with speeds comparable to those inferred from observations.
Furthermore, we find that the UV/optical fluxes of our model are compatible with a damped-random-walk (DRW) process, with a damping time $\taud$ consistent with observations.
The scaling relation between $\taud$ and the rest-frame wavelength $\lambda$ has a bended shape, $\taud\propto\lambda$ at short wavelengths and transitioning to a plateau at long wavelengths.
At $\lambda=2500\text{\AA}$,
the damping time roughly follows $\propto\MBH^{1/2}$ when $\MBH\gtrsim 10^6\Msun$, consistent with observational constraints, though it tends to be underestimated for lower SMBH masses.
Including additional refinements, such as the dependence of dynamo properties on $M_\text{BH}$ and AGN luminosity, and accounting for X-ray reprocessing, would further enhance the accuracy of the model.
In addition, we show that generic disk models with spatially uncorrelated fluctuations cannot explain the observed DRW damping times; spatially correlated fluctuations, such as those discussed in this paper, may be an essential ingredient.

\end{abstract}

\section{Introduction}

Accretion flows around supermassive black holes (SMBHs) at galaxy centers provide one of the most compelling laboratories for studying plasmas under extreme physical conditions.
It has long been established that the UV/optical emissions from active galactic nuclei (AGNs) exhibit stochastic variability \citep[e.g.,][]{Ulrich+1997,Giveon+1999,Kelly+2009, MacLeod+2010, Burke+2021,Tang+2023}, characterized by several peculiar features whose origins remain elusive.
Since the UV/optical band coincides with the thermal emission from a thin accretion disk surrounding a SMBH, understanding its variability is crucial as it may encode vital information about the accretion disk.

A possible origin of the disk emission variability is reverberation,
in which a lamppost-like hot corona illuminates the disk and drives the variability \citep{Krolik+1991,Shappee+2014,Fausnaugh+2016,Edelson+2017,Secunda+2024}.
The reverberation model naturally links X-ray and UV/optical variabilities through light-travel time delays, allowing for direct constraints on disk geometry.

Alternatively, \cite{NeustadtKochanek2022} developed a method to map AGN continuum light curves to time-dependent radial profiles of disk temperature, assuming small temperature perturbations over a Shakura-Sunyaev profile \citep{ShakuraSunyaev1973}.
The method has been applied to seven AGNs using AGN STORM data, and subsequently to a sample of SDSS quasars \citep{StoneShen2023} and Mrk 817 \citep{Neustadt+2024}.
These works have consistently revealed coherent temperature fluctuations that move both inward and outward over the disk,
with a relative amplitude of $2$-$10\%$.
More importantly, the radial speed of such fluctuations is slow, at the order of $0.01$-$0.1c$, and incompatible with a reverberation origin.

The slowly-moving temperature fluctuations suggest that some long-time UV/optical variability may be intrinsic to the disk, possibly from turbulence driven by the magnetorotational instability \citep[MRI;][]{King+2004,MayerPringle2006,JaniukCzerny2007,HoggReynolds2016}.
%So far, most models of variable disks typically focus on the resulting disk emission,
%but the disk fluctuations or the source perturbing them have remained to be assumed without a physical origin
%\citep{Lyubarskii1997,DexterAgol2011,CowperthwaiteReynolds2014,Cai+2016,
%Sun+2020,TurnerReynolds2021}.
In \cite{Kaul+2025}, the authors performed radiative magnetohydrodynamical simulations and argue that fast magnetosonic waves in the disk may drive slowly moving temperature fluctuations when the Maxwell stress is dominated by the turbulent component.

In this work, we propose a new mechanism to drive slow disk surface temperature fluctuations, namely by quasi-periodic large-scale dynamos (LSDs).
Local and global simulations have revealed that LSDs driven by MRI in thin disks exhibit outgoing dynamo waves with typical length scales of the radius $r$ and periods $\CO\sim10$ times the orbital time
\citep[e.g.,][]{Gressel2010,BaiStone2013,HoggReynolds2018}.
Dynamo waves then travel at a speed of $\sim\CO^{-1}$ times the Keplerian speed, comparable to the local sound speed for a thin disk.
Our model differs from many previous studies on disk perturbations,
in that we provide a physical mechanism of driving stress fluctuations in the disk,
rather than prescribing the statistics of the perturbations.
Our model is semi-analytical with a few parameters that describe the nature of the LSD, providing a potential way of extracting unobservable dynamo parameters from disk emissions.

Furthermore, our model predicts disk light curves that are compatible with damped-random-walk (DRW) processes with plausible damping times.
Although the DRW feature is not universal for AGNs
\citep[e.g.,][]{Mushotzky+2011,Gonzalez-MartinVaughan2012,Edelson+2014,Caplar+2017,Smith+2018,Arevalo+2023,Arevalo+2024,Petrecca+2024},
it has been reported that a fraction of the AGN UV/optical light curves agrees with the DRW model \citep{Kelly+2009, MacLeod+2010,Gonzalez-MartinVaughan2012,Zu+2013,Burke+2021,Helias+2025}.
The typical damping time $\taud$ has been found to depend weakly on the rest-frame luminosity or the redshift, mildly on the wave length $\lambda$,
but more strongly on the SMBH mass $\MBH$ with a power-law index $\lesssim 0.5$-$1$ \citep{Kelly+2009, MacLeod+2010, Burke+2021, Wang+2023, Arevalo+2024,Helias+2025}.
The $\taud-\MBH$ relation is sufficiently tight to be inverted,
allowing for SMBH mass estimates based on $\taud$ measurements
that are consistent with those obtained from virial methods \citep{Rachana+2025}.

Observations beyond AGNs and in other wavebands offer some hints
regarding the elusive origin of the observed DRW-like UV/optical variability and the $\taud(\MBH,\lambda)$ scalings.
The $\taud\sim \MBH^{0.5}$ scaling has been found to apply to accreting stellar-mass black holes and white dwarfs, as well as young stellar objects \citep{Scaringi+2015,Su+2024}, indicating that the scaling is likely universal for accretion disks and does not have a general-relativistic origin.
%, such as the Bardeen-Petterson effect \citep{BardeenPetterson1975, NelsonPapaloizou2000}.
Furthermore, similar scalings have been observed in the X-ray band \citep{Zhang+2024}, and correlations among $\taud$, the jet magnetic field, and the black hole spin have been found in jetted AGNs \citep{Chen+2025}.
These observations provide indirect evidence that DRW-type variabilities may be closely linked to magnetic processes in accretion flows.

The rest of the paper is organized as follows.
In Section~\ref{sec:eqns}, we introduce our fluctuating disk model and the governing equations.
In Section~\ref{sec:results}, we showcase and explain the numerical results for the variabilities produced by our model.
We conclude in Section~\ref{sec:conclusion}.

\section{Dynamo-driven fluctuating disk model and governing equations}
\label{sec:eqns}

We consider a geometrically thin, optically thick disk with a Keplerian rotation profile $\Omega(r)$.
Assuming that the angular momentum transport is dominated by the turbulent viscous stress, the vertically integrated and azimuthally averaged viscous-diffusive equation of the surface density $\Sigma$ is \citep[see, e.g.,][]{Frank+2002}
\beq
\partial_t\Sigma-\frac{3}{r}\partial_r\left[
r^{1/2}\partial_r\left(r^{1/2}\nu\Sigma\right)
\right]=0,
\label{eqn:disk_eqn}
\eeq
where $\nu(t,r)$ is the turbulent viscosity with spatial and temporal variabilities.
At radius $r$, the accretion rate is given by
\beq
\Mdot(t,r)
%=-2\pi r\Sigma v_r
=6\pi r^{1/2}\partial_r\left(r^{1/2}\nu\Sigma\right),
\label{eqn:Mdot}
\eeq
and the energy dissipation rate per unit surface area is
\beq
D(t,r)=\frac{1}{2}\nu\Sigma(r\partial_r\Omega)^2=\frac{9}{8}\nu\Sigma\Omega^2.
\eeq
For an optically thick disk, the radiating blackbody temperature at the disk surface is $T(t,r)=\left(D/\sigma_\text{SB}\right)^{1/4}$,
with $\sigma_\text{SB}$ being the Stefan-Boltzmann constant.
The specific flux at photon wavelength $\lambda$ from a face-on disk at distance $d$ is then
\beq
F_\lambda =\frac{4\pi hc^2}{d^2\lambda^5}\int_{\rin}^{\rout}
\frac{r\ \text{d}r}{e^{hc/\lambda k_\text{B}T}-1},
\label{eqn:Fnu}
\eeq
where $h$ is the Planck constant, $c$ is the speed of the light, $k_\text{B}$ is the Boltzmann constant, and $r_0$ and $\rout$ are the inner and outer radii of the disk, respectively.
The total luminosity from the thermal emission is
\beq
L(t)=4\pi\int_{\rin}^{\rout} D(t,r)\ r\text{d}r.
\label{eqn:L}
\eeq

The accretion rate, disk spectrum, and total luminosity can be calculated from Equations~(\ref{eqn:Mdot})-(\ref{eqn:L}) once the turbulent viscosity is given, which we detail in the next subsection.

\subsection{Description of turbulent viscosity}

We consider a Shakura-Sunyaev type viscosity that is scaled by the local sound speed $\cs(r)$ and disk scale height $H(r)$,
both of which vary smoothly over the disk according to a steady-state solution.
However, the proportional dimensionless parameter $\alpha$ is assumed to have both spatial and temporal variabilities due to a disk dynamo, so that
\beq
\nu(t,r)=\alpha(t,r)\cs(r) H(r).
\label{eqn:nu0}
\eeq
The vertical force-balance condition $\cs=\Omega H$ together with a dimensionless scale height $\epsilon=H/r\propto r^{1/8}$ due to Kramers' law for opacity \citep{Frank+2002} gives
\beq
\nu(t,r)=\alpha(t,r)\cso H_0\tilde r^{3/4},
\label{eqn:nu}
\eeq
where the subscript $0$ indicates quantities evaluated at the disk inner boundary $r=r_0$, and $\tilde r=r/r_0$.
The relation $\cs^2\propto T\propto \nu^{1/4}\propto \alpha^{1/4}\cs^{1/2}$ yields the weak dependence $\cs,H\propto \alpha^{1/6}$, which ensures that $\cs$ and $H$ vary much weaker than $\alpha$ and can be approximated by steady-state solutions.

We assume that the disk angular momentum transport is provided by the MRI turbulence.
The dimensionless viscosity parameter $\alpha$ is defined using the mean of the turbulent Reynolds and Maxwell stresses, i.e.,
\beq
\alpha=\frac{\rho\left(\bar{u_ru_\phi}-\bar{b_rb_\phi}\right)}{P},
\label{eqn:def_alpha}
\eeq
where $\rho$ is the gas density, $P$ is the thermal pressure,
and $\bm u$ and $\bm b$ are the turbulent velocity and magnetic field, respectively.
The magnetic field is measured in the Alfv\'en unit,
and the overlines indicate azimuthally averaged quantities.
We model the variability of $\alpha$ by assuming that
(i) the turbulent Maxwell stress dominates in Equation~(\ref{eqn:def_alpha}), as has been routinely observed in simulations \citep[see][for a summary]{Blackman+2008},
and
(ii) the turbulent Maxwell stress has a stationary part representing a steady-state turbulence, and a variable part because of its response to a quasi-periodic LSD.
Correspondingly we write
\beq
\alpha(t,r)=\alpha_0\left[1+\alphaLSD(t,r)\right],
\eeq
where $\alpha_0$ is the stationary part of the viscosity parameter,
and $\alphaLSD$ is the normalized variable part.
While the turbulence time scale is $\sim\Omega^{-1}$ and its variability is averaged out when considering accretion dynamics,
the LSD time scale is $\sim10$ times longer \citep{Gressel2010, BaiStone2013, HoggReynolds2016} and may leave an imprint on the accretion rate during the observation interval.
For example, at $10$ gravitational radii away from a  $10^8\Msun$ black hole, the turbulence and LSD time scales are $0.2$ day and $2$ days respectively, while the observation time scale can be of hundreds of days with a cadence of days.

We consider the variable part $\alphaLSD(t,r)$ to be driven by the quasi-periodic LSD waves,
which we detail in the next subsection.

\subsection{Prescription of dynamo waves}

\begin{figure*}
\centering
\includegraphics[width=0.9\textwidth]{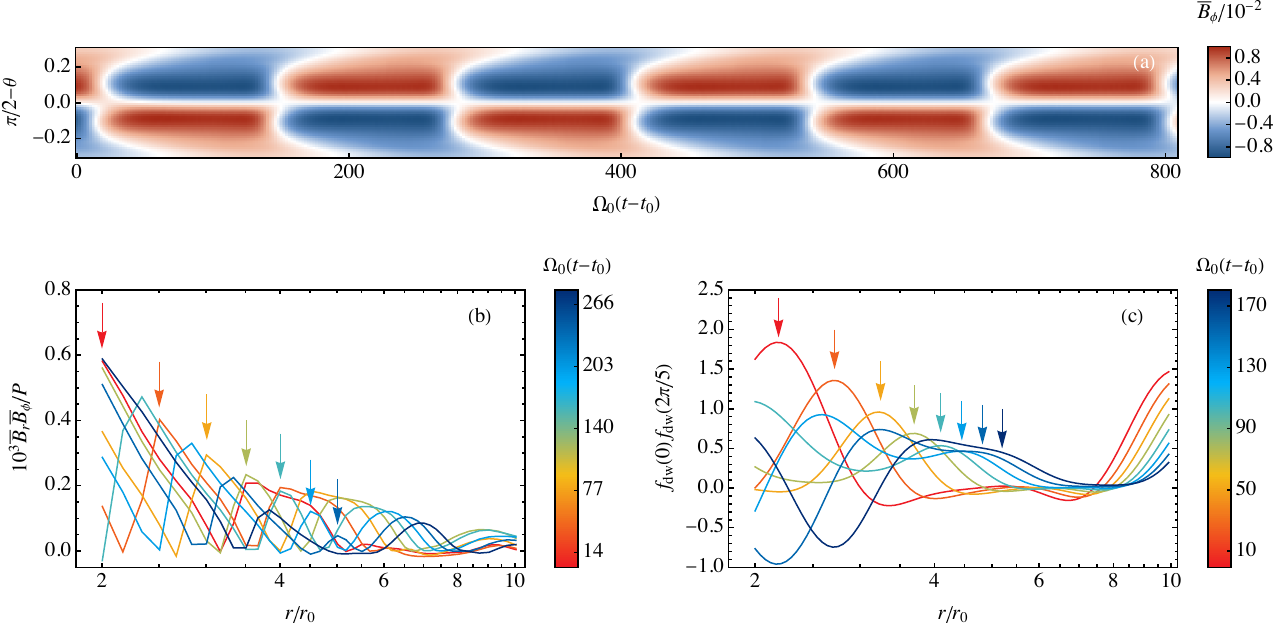}
\caption{Comparing our prescription Equation~(\ref{eqn:alphass(B)}) with the global disk dynamo simulation in \cite{Zhou2024}.
Panels (a) and (b) are outputs of run~\texttt{AO1} in \cite{Zhou2024}.
Panel (a) shows the space-time diagram of $\bar B_\phi(t,\theta)$ at $r=2\rg$, starting from $t_0=10^4\Omega_0^{-1}$, and $\theta$ is the latitude.
Panel (b) plots the normalized Maxwell stress at $\theta=0.15$, with $t_0=5960\Omega_0^{-1}$.
Panel (c) shows the varying part in Equation~(\ref{eqn:alphass(B)}) using the parameters $\alpha=0.3$, $\epsilon_0=0.1$, $\Cl=30$, and $\CO=20$, starting from $t_0=2\times10^6\Omega_0^{-1}$.
The arrows in panels (b) and (c) remark a particular peak in each panel which propagates, damps, and merges with a later peak.}
\label{fig:model_compare}
\end{figure*}

LSDs that amplify disk-scale magnetic fields can be realized in high-resolution simulations that resolve a sufficiently extended turbulent inertial range \citep{Gressel2010,BaiStone2013,HoggReynolds2018,Liska+2020,Dhang+2020}, or low-resolution simulations which incorporates sub-grid dynamo terms \citep{vonRekowski+2003,BucciantiniDelZanna2013,Bugli+2014,StepanovsFendt2014,Skadowski+2015,FendtGassmann2018,Dyda+2018,Tomei+2020,VourellisFendt2021, Zhou2024,Zhou+2025}.
For thin disks, both local and global simulations display dipolar field configurations and radially outgoing dynamo waves.
The coherent length is $\sim30$ times and the time scale is $\sim10$
times that of the turbulence.
Figure~\ref{fig:model_compare}(a) gives an example of the space-time diagram of the LSD-amplified large-scale magnetic field $\bar B_\phi$ using the data of run~\texttt{AO1} in \cite{Zhou2024}.
Instead of directly using simulation data to calculate $\alphaLSD$ or the variability,
we consider a semi-analytical prescription of the mean magnetic field $\bar B_{r,\phi}(t,r)$ in the disk.
The model resembles the key feature of previous disk or shearing-box simulation that LSD generates dynamo waves,
and is more flexible and affordable for parameter surveys or fitting.

We consider only the radial variation of $\bar B_r\bar B_\phi$,
because
(i) due to the dipolar LSD geometry,
the mean-field stress $\bar B_r\bar B_\phi$ have a uniform sign across the disk mid-plane, and
(ii) we consider vertically averaged dynamics.
At radius $r$, the dynamo wave has a local frequency $\omegaD$ and wavenumber $\kD$,
\beq
\omegaD(r)=\frac{\Omega(r)}{\CO},\ 
\kD(r)=\frac{2\pi}{\Cl H(r)},
\eeq
where $\CO$ and $\Cl$ are dimensionless numbers to parameterize the LSD scales.
A plausible ansatz for the outgoing dynamo waves is
\beq
\bar B_{r,\phi}(t,r)\propto \sin\left[
\omegaD(r) t-\kD(r) r
\right],
\eeq
i.e., the magnetic fields propagate at the local frequency and wavenumber.
However, this approach is problematic both numerically and physically:
The radial dependence of $\omegaD$ causes the magnetic fields at adjacent radial locations to gradually fall out of phase over time, and eventually give arise to spatial variations over arbitrarily small scales.
In reality, LSDs cannot produce structures smaller than the turbulence scale due to turbulent diffusion.
Instead, the dynamo waves are observed in simulations to be modulated into several wave packets, each having roughly constant wave frequency, with smooth boundaries mediated by turbulent diffusion.
At such boundaries, the adjacent dynamo wave peaks will merge,
as demonstrated in Figure~\ref{fig:model_compare}(b) using the data from run~\texttt{AO1} of \cite{Zhou2024}.
Therefore, a more accurate representation of disk dynamo patterns would involve the superposition of multiple wave packets, each centered at a different radius.

We assume that the disk LSD is active in the radial range $\xiin\leq r\leq\xiout$,
and $\xiout$ is chosen to be smaller than $\rout$ to allow for a stable boundary condition.
To construct dynamo wave packets, we assume that the $i$-th dynamo wave packet is centered at the radial location $\xi_i$, and is damped using the Gaussian profile over a distance equal to one dynamo wavelength, $2\pi/\kD(\xi_i)$.
The distance between two neighboring dynamo sites, $\xi_i$ and $\xi_{i+1}$, is then $2\pi/\kD(\xi_i)\propto \xi_i^{9/8}$, resulting in logarithmically distanced dynamo wave packets across the disk.

The superposition of all the dynamo waves is then
\begin{align}
f_\text{LSD}(\Delta\varphi)=&\sum_i \exp\left[-\frac{(r-\xi_i)^2}{2\Delta\xi_i^2}\right]\notag\\
&\times\sin\left[\omegaD(\xi_i) t-\kD(\xi_i) r+\varphi_i+\Delta\varphi\right],
\end{align}
where we have introduced the initial phase of each wave mode, $\varphi_i$,
and an overall phase, $\Delta\varphi$.
To avoid the initial coherence among different wave components,
we use $\varphi_i=2\pi(i-1)/(n_\text{dyn}-1)$,
where $n_\text{dyn}$ denotes the total number of dynamo wave components and can be calculated once $\kD$ and $\xiout$ is given.
The particular form of $\varphi_i$ is used for the sake of reproducibility,
and using random initial phases does not qualitatively change our results.
An overall phase difference of $\Delta\varphi=2\pi/5$ between $\bar B_r$ and $\bar B_\phi$ is empirically extracted from the simulations \citep{Gressel2010,Zhou2024},
so that $\bar B_r\propto f_\text{LSD}(0)$ and $\bar B_\phi\propto f_\text{LSD}(2\pi/5)$.

Finally, we assume that the variable part of the turbulent Maxwell stress response linearly to the mean stress $\bar B_r\bar B_\phi$, and hence
\beq
\alpha=\alpha_0\left[1+C_\beta f_\text{LSD}(0)f_\text{LSD}\left(\frac{2\pi}{5}\right)\right],
\label{eqn:alphass(B)}
\eeq
where $C_\beta$ is roughly equal to the time average of $\bar B_r\bar B_\phi/\bar{b_rb_\phi}$,
a scaling factor accounting for the relative strengths between the large- and small-scale magnetic fields.
In panels (b) and (c) of Figure~\ref{fig:model_compare}, we compare the evolution of the normalized Maxwell stress $\bar B_r\bar B_\phi/P$ from run~\texttt{AO1} of \cite{Zhou2024} to the corresponding dimensionless factor $f_\text{LSD}(0)f_\text{LSD}\left(\frac{2\pi}{5}\right)$ in Equation~(\ref{eqn:alphass(B)}).
We adopt the disk and dynamo coefficients that are comparable to \cite{Zhou2024}: $\alpha_0=0.3$, dimensionless scale height $\epsilon_0=0.1$, $\CO=20$ and $\Cl=30$.
As indicated by the arrows in the two panels, our semi-analytical model captures both the radially outgoing dynamo waves and the merging between two adjacent waves that is mediated by turbulent diffusion.
We also note that we do not aim to exactly reproduce the field patterns as seen in simulations, but merely to build a model that captures the essential variability of a LSD, namely that superposition of outgoing dynamo waves.

Note that in Equation~(\ref{eqn:alphass(B)}),
the variable part can be transiently negative [e.g., at $r\simeq 7r_0$ in Figure~\ref{fig:model_compare}(b) and $r\simeq2.5r_0$ in Figure~\ref{fig:model_compare}(c) for the shown time intervals],
indicating that, temporally, the mean-field contribution to the Maxwell stress can be opposite to that of the turbulent contribution.
However, non-local angular momentum transport may also occur due to the presence of mean fields \citep{BlackmanNauman2015}, e.g., through disk winds, which always dissipates the disk angular momentum regardless of the direction of the mean fields.
To account for such effects, we set a floor value of $\alpha$ to be $0.1\alpha_0$, so that the turbulent viscosity is always positive.

\subsection{Fiducial parameters and unit conversion}

Equations~(\ref{eqn:disk_eqn}), (\ref{eqn:nu}) and (\ref{eqn:alphass(B)}) are the governing equations, which we implement and numerically solve in the publicly available \texttt{Pencil Code} \citep{JOSS2021} using sixth-order accurate finite differences and the third-order Runge-Kutta time-stepping scheme.
Except for the thermal emission part, the one-dimensional diffusion equation is scale-free,
and the disk dynamics is governed by the following fiducial dimensionless parameters,
\beq
\rout/r_0=120,\ 
\epsilon_0=0.1,
\alpha_0=0.1.
\eeq
The fiducial dimensionless dynamo parameters are
\beq
\Cbeta=0.3,\ \CO=8,\ \Cl=30,\ \xiout/r_0=100.
\eeq

We use the initial condition $\Sigma(t=0,r)=\Sigma_0=10^{-5}$ in code units,
and the same value is used as the density floor to avoid negative values of the surface density.
At both the inner and the outer radial boundaries, $\Sigma$ is kept fixed, with $\Sigma(r_0)=\Sigma_0$ as a sink and $\Sigma(\rout)=10^3\Sigma_0$ as a mass source.
The very small value $\Sigma_0$ at $r_0$ ensures that mass can only diffuse radially inward at the inner boundary.

To convert code units to physical units, we adopt the fiducial length scale and black-hole mass
\beq
r_0=10\rg=\frac{10G\MBH}{c^2},\ 
\MBH=10^8\ \Msun.
\eeq
Using the chosen inner disk radius,
we do not consider any general relativistic effect or variable source relevant for the X-ray variability \citep{Chartas+2016,Dogruel+2020}, such as instability or magnetic reconnection that could happen in a jet.
The accretion rate at the outer boundary, calculated using Equation~(\ref{eqn:Mdot}), is used to scale $\Sigma$ to physical units by setting the mean Eddington ratio $\etaEdd=L/L_\text{Edd}$ in the steady state to the fiducial value $0.2$, which also fixes the temperature unit conversion.
To convert to the observed flux, we adopt a distance of $d=100$ Mpc.

\section{Results}
\label{sec:results}

In this section, we first demonstrate some of the persistent features of our model, namely the DRW-like power-spectral densities (PSDs) and the scaling relations of $\taud$ with photon wavelengths and SMBH masses.
We then investigate how $\taud$ depends on the disk and dynamo parameters, including $\Cbeta$, $\CO$, and $\alpha$.
We also make a comparison with the model of \cite{Lyubarskii1997} at the end of this section.

\subsection{Temperature fluctuations}

\begin{figure}
\centering
\includegraphics[width=0.48\textwidth]{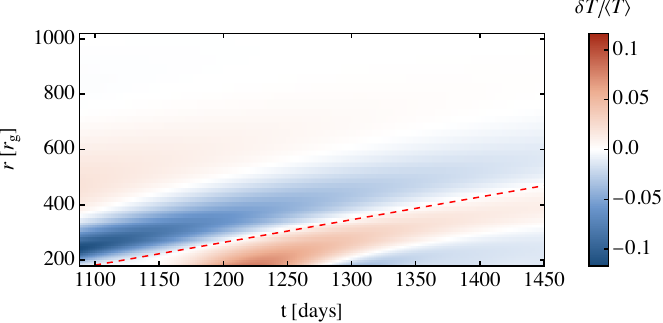}
\caption{Relative temperature fluctuations for the fiducial run.
The dashed curve indicates the trajectory of a patch moving at the local dynamo wave speed.}
\label{fig:deltaT}
\end{figure}

In Figure~\ref{fig:deltaT} we plot the relative temperature fluctuation $\delta T/\abra{T}$ (with $\abra{T}$ being a time average) in a representative time interval.
The amplitude of the fluctuations is stronger at the inner disk, and reaches $<10\%$ level in the middle to outer disk,
in broad agreement with observations where $\delta T/T\sim 2-4\%$
\citep{NeustadtKochanek2022,StoneShen2023,Neustadt+2024}.
The fluctuations move radially outward, as it is dominated by the dynamo waves rather than the slower viscous accretion.
Given the fiducial parameters, the local wave speed is comparable to the sound speed,
\beq
\frac{v_\text{LSD}}{\cs}=\frac{\omegaD/\kD}{\cs}=\frac{\Cl}{2\pi\CO}\simeq0.6,
\eeq
or $\simeq 0.008$ times the speed of light.
A trajectory following $\dot r=v_\text{LSD}$ is shown in the dashed curve in Figure~\ref{fig:deltaT}, matching the temperature pattern well.

\subsection{PSDs of accretion rates}

We next examine the variability of the accretion rate $\Mdot$ at the inner disk boundary and its PSD.
The accretion rate directly relates to the total energy  output in the disk,
and is a fundamental aspect that applies regardless of the disk emission mechanisms.
By investigating the accretion variability, we can gain insights that may also be applicable to studies involving optically thin disks, where the emission mechanisms differ but the underlying energy source remains the same.

\begin{figure*}
\centering
\includegraphics[width=0.9\textwidth]{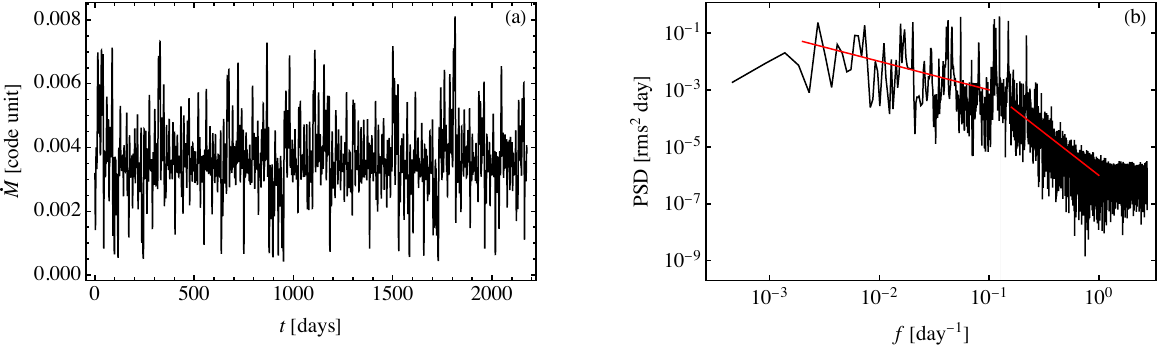}
\caption{Results for the fiducial run:
(a) The time series of accretion rate at the inner boundary.
(b) The corresponding PSD.
In the right panel, the red lines indicate $\propto f^{-1}$ (left) and $\propto f^{-3}$ (right), respectively.
The vertical line indicates the local dynamo frequency, $f=\Omega_0/\CO$.}
\label{fig:fid_weak_b/Mdot_ts_psd}
\end{figure*}

As shown in Figure~\ref{fig:fid_weak_b/Mdot_ts_psd}, $\Mdot$ exhibits large variability and a clearly bended PSD.
A low frequencies, the power index of the PSD is marginally consistent with $f^{-1}$,
as predicted by the fluctuation-propagation model of \cite{Lyubarskii1997}.
The power law is predicted to extend up to the local variability frequency,
$\Omega_0/\CO$, as marked by the vertical line in the plot.
At high frequencies, the PSD becomes close to $\propto f^{-3}$.
Overall, both low- and high-frequency slopes are steeper than those reported by \cite{TurnerReynolds2021}.
The difference of the driven $\alpha$ fluctuations between our model and theirs lies in that
(i) temporally, ours is quasi-periodic and theirs is stochastic, and
(ii) spatially, ours is coherent over $30H$ and theirs is over $H$.
Both signatures may contribute to the difference seen in the PSDs.

When getting to close the black hole, the interaction between a thin disk, the black hole, and potentially a thick disk and jets becomes non-negligible.
The main target of this work is to provide a mechanism to drive the UV/optical variability intrinsic to the disk,
but here we also postulate that when reaching the inner disk region,
such fluctuations are able to modulate or provide the needed source of variabilities in other bands originating from the innermost accretion region, including:
(i) high-energy X-rays or $\gamma$ rays \citep{Zhang+2024} which may come from the innermost disk or a compact hot corona,
(ii) sub-millimeter bands \citep{Chen+2023} caused by jets,
and (iii) a source of X-ray variability in reverberation models \cite[e.g.,][]{Sun+2020,Hagen+2024}.
In the last scenario, the disk variability has both intrinsic contributions  and a reprocessed and reverberated part due to X-rays,
and both should be taken into account in a future variability model.

\subsection{Bolometric luminosity}

\begin{figure*}
\centering
\includegraphics[width=1\textwidth]{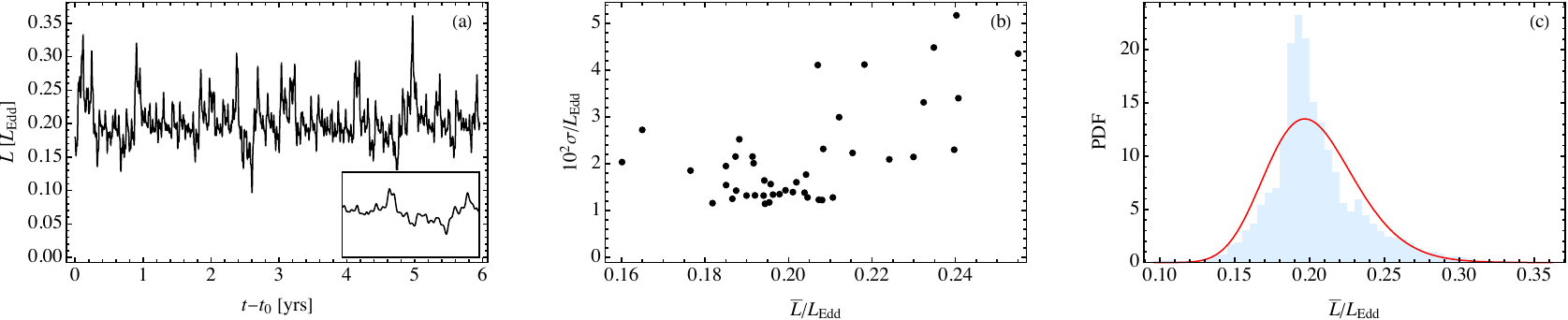}
\caption{Results for the fiducial run.
(a) The bolometric luminosity normalized by the Eddington luminosity $\LEdd$,
and the inset is a zoom-in plot for the $t-t_0=800-1000$ days interval on the same vertical scale.
(b) The rms-flux correlation for the binned flux.
(c) The PDF of the flux and a log-normal fit (red curve).
}
\label{fig:rms_flux}
\end{figure*}

In Figure~\ref{fig:rms_flux}, we plot the bolometric luminosity of the fiducial run in the steady state calculated from Equation~(\ref{eqn:L}).
Following the method of \cite{Smith+2018},
we bin the light curve over $2$ days and then cut into $50$-day intervals.
For each interval, we calculate the mean and the standard deviation,
as shown in panel~(b).
The lack of a clear linear rms-flux relation suggests that our model does not exhibit strict multiplicativity that would be expected from a fluctuation-propagation model \citep{UttleyMcHardy2001,Uttley+2005}.
Additionally, panel~(c) shows the probability distribution function (PDF) of the flux which cannot be well-fitted by a log-normal curve, again deviating from the expectation of multiplicative fluctuations.
We note that \textit{Kepler} observations do not report a linear rms-flux relation \citep{Smith+2018}.
As for the flux PDF, over time scales of years, the flux differences from individual quasars follow log-normal distributions \cite[e.g.,][]{MacLeod+2012}, but probably not over shorter time scales \citep{KozlowskiSzymon2016,Smith+2018} where bi-modal or normal distribution is also seen.
Both features are in broad agreement with our model.

Generally, a possible cause of the deviation from multiplicative fluctuations could be the coherence of the fluctuation driver at different radii, including irradiation and outflow-driven perturbations.
In our case, the outward-propagating dynamo waves can counter inward viscous fluctuations, partially intercept the multiplicative behavior.
When fluctuations are driven independently \citep{CowperthwaiteReynolds2014} or coherently over small distance \citep{TurnerReynolds2021}, both the linear rms-flux relation and a log-normal PDF of flux can be recovered.

\subsection{Specific fluxes and damping times}

\begin{figure*}
\centering
\includegraphics[width=0.9\textwidth]{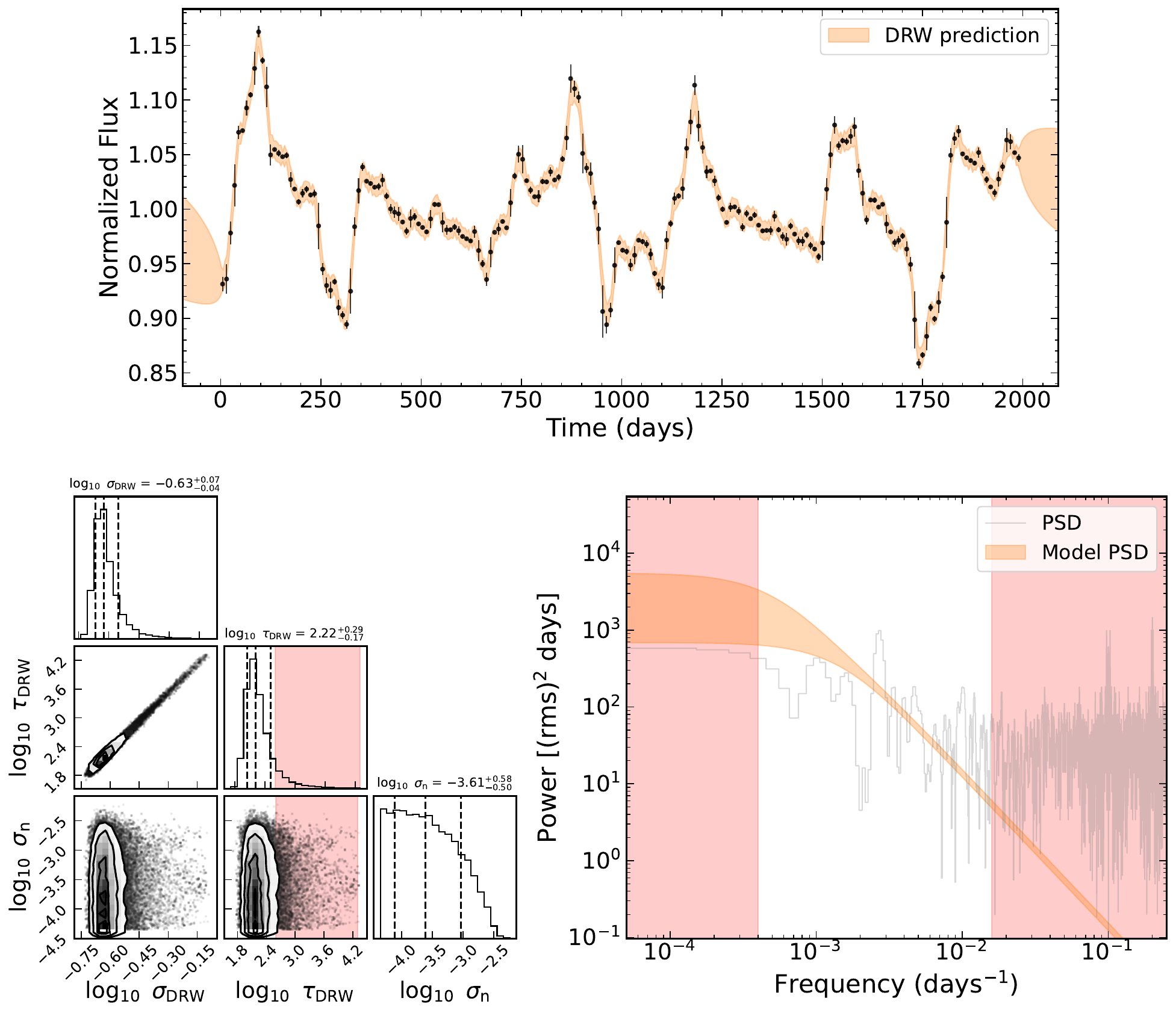}
\caption{DRW-model fitting for the fiducial run.
Top panel:
The normalized specific flux at $2500\text{\AA}$ binned with a $10$-day interval, and its DRW prediction.
Bottom-left panel:
Corner plot of the fitted DRW amplitude $\sigma_\text{DRW}$, damping time $\tau_\text{DRW}$, and white-nose term $\sigma_\text{n}$.
The vertical dashed lines indicate the $16$th, $50$th, and $84$th percentiles, respectively.
Bottom-right panel:
The PSD and the fitted DRW model.
In the top and bottom-right panels,
the orange shaded areas demark the $1\sigma$ uncertainty.
In the bottom panels,
the red shaded regions mark time scales greater than $20\%$ of the light curve length and less than the mean cadence.}
\label{fig:fid_tau_fit}
\end{figure*}

We now examine the specific fluxes $F_\lambda$ of the fiducial run calculated using Equation~(\ref{eqn:Fnu}).
We use the \texttt{taufit}%
\footnote{https://github.com/burke86/taufit/tree/master} package \citep{Burke+2021}, which provides DRW-model fitting based on the fast Gaussian process solver \texttt{Celerite} \citep{Foreman-Mackey+2017}.
For the fiducial run, the normalized specific flux at $\lambda=2500\text{\AA}$,
the fitted parameters, and the fit of the PSD are shown in Figure~\ref{fig:fid_tau_fit}.
The PSD resembles DRW-like shapes at low to intermediate frequencies, and flatten out at high frequencies \citep{Mushotzky+2011,Zu+2013}.
The damping time is found to be $\taud= 164^{+157}_{-53}$ days, longer but consistent with observations \citep{Burke+2021,Zhang+2024,Helias+2025}.
We ensure that our baseline is sufficiently long to avoid underestimating the damping time \citep{Sun+2020,Stone+2022,Zhou+2024,Ren+2024}.

It is worth noting that the input variability in Equation~(\ref{eqn:alphass(B)}) is not prescribed to follow the DRW process; rather, the PSD of $\alpha$ at a fixed radius approximately exhibits a single peak corresponding to the local dynamo frequency.
Interestingly, the overall disk dynamics naturally evolve to produce a DRW-like PSD.
As we will discuss in Section~\ref{sec:L97}, implementing the \cite{Lyubarskii1997} model using Poisson processes cannot reproduce the DRW-shaped PSD;
or it yields too long a damping time incompatible with observations.
A more detailed comparison is deferred to that subsection.

\begin{figure}
\centering
\includegraphics[width=0.45\textwidth]{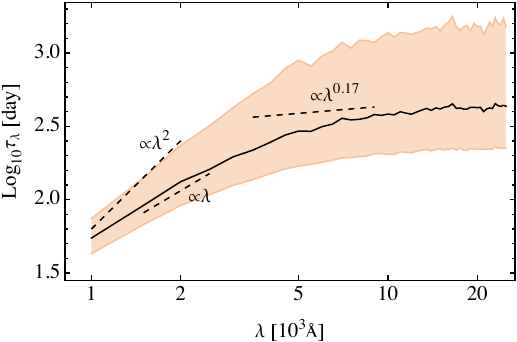}
\caption{For the fiducial run, the $\lambda$ dependence of the damping time scale.
The colored region indicates the $1\sigma$ uncertainty.
Three power-law relations are shown for reference.}
\label{fig:r120b/tau-lambda}
\end{figure}

To examine how the damping time scale depends on the emitting wavelength,
the same fitting procedure is done for a wider range of wavelengths,
and the fitted values of $\taud$ are shown in Figure~\ref{fig:r120b/tau-lambda}.
A dashed line indicating a $\lambda^2$ proportionality is shown for reference,
as it will be expected if photons at wavelength $\lambda$ predominantly originate from the ring around $r_\lambda$ where the thermal spectrum peaks at $\lambda$ \citep[see][and also Section~\ref{sec:conclusion}]{MacLeod+2010}.
The calculated $\lambda$ dependence is closer to $\propto\lambda$ and flattens out because the emission from lower-energy photons arises not only from the ring at $r_\lambda$ but also from regions at $r<r_\lambda$, leading to a blending of their respective time scales.
\citet{MacLeod+2010} found a weak scaling $\taud\propto \lambda^{0.17}$ from $9275$ quasars in SDSS Stripe 82 with light curves in all the five passbands, which we show for reference in the $[3500,9000]\text{\AA}$ waveband in Figure~\ref{fig:r120b/tau-lambda}.
Note that our $\taud$ calculation is based on the fiducial values $\MBH=10^8\Msun$ and $\etaEdd=0.2$, but can be readily rescaled:
Since the disk surface temperature scales as $T^4\propto\nu\Sigma\Omega^2\propto\MBH^2\etaEdd$,
some typical wavelength respects $\lambda\propto T^{-1/4}\propto\MBH^{-1/2}\etaEdd^{-1/4}$.
Hence the critical wavelength at which the $\taud-\lambda$ relation flattens will scale with the black hole mass and the Eddington ratio as
\beq
\lambda_\text{flat}
\simeq5000\text{\AA}
\left(\frac{\MBH}{10^8\Msun}\right)^{-1/2}
\left(\frac{\etaEdd}{0.2}\right)^{-1/4}.
\eeq
Mixing samples with different SMBH masses and Eddington ratios will likely obscure the underlying $\taud-\lambda$ relation, since the latter depends explicitly on both $\MBH$ and $\etaEdd$.

%In Figure~\ref{fig:r120b/tau-lambda}(b), we plot the $\lambda$ dependence of the PSD slope.
%Although broadly consistent with a $-2$ slope, the data shows a systematic trend:
%$p_\lambda\simeq-2$ only in the window $[2000,5000]\text{\AA}$,
%and $|p_\lambda|$ increases with decreasing $\lambda$.
%Our model indicates that the PSD becomes steeper at shorter wavelengths,
%and together with Equation~(\ref{eqn:lambda_T}),
%implies that the PSD at some fixed color becomes steeper for lower SMBH mass, which is consistent with $\textit{Kepler}$ data \citep{Smith+2018}. (refs).

\begin{figure}
\centering
\includegraphics[width=0.45\textwidth]{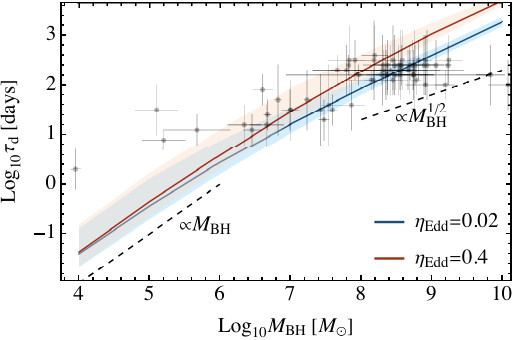}
\caption{For the fiducial run, damping time $\taud$ at $2500\text{\AA}$ versus $\MBH$.
The colored regions indicate the $1\sigma$ uncertainty for each curve.
The gray points with error bars represent observational data from Figure 1 of \cite{Burke+2021} for reference.}
\label{fig:tau2500_MBH}
\end{figure}

Finally, we compute $\taud$ at $2500\text{\AA}$ for $\MBH\in[10^4,10^{10}]\Msun$ with $\etaEdd=0.02,0.4$.
For each run, the light curve is binned using a time scale based on the SMBH mass, as
\beq
t_\text{bin}=10\text{ days}\times\left(\frac{\MBH}{10^8\Msun}\right)^{3/2}.
\eeq
In Figure~\ref{fig:tau2500_MBH}, we plot the two $\taud-\MBH$ relations with the chosen values of $\etaEdd$ in blue and red, respectively,
along with the data from Figure~1 of \cite{Burke+2021} in gray data points with error bars.
In our model, the $\taud-\MBH$ scaling law varies when $\etaEdd$ is changed:
At $\MBH\lesssim 10^6\Msun$, the scaling $\taud\propto\MBH$ is roughly independent of $\etaEdd$, while at larger $\MBH$, the dependence on $\etaEdd$ is slightly stronger, introducing a weaker scaling between $\taud$ and $\MBH$ when $\etaEdd$ is small.
In the $\MBH\gtrsim10^6\Msun$ mass range, the scaling $\taud\sim \MBH^{1/2}$ is in agreement with the flux-weighted average of the orbital time scale \citep{Wolf+2024},
and reasonably well covers the observational data given the uncertainties in $\etaEdd$, whereas at lower $\MBH$ the scaling deviates.
The deviation could be due to an additional dependence of the dynamo properties on the SMBH mass or accretion rates,
as explored in the next subsection.

\subsection{Parameter survey}
\label{sec:param}

\begin{figure*}
\centering
\includegraphics[width=0.75\textwidth]{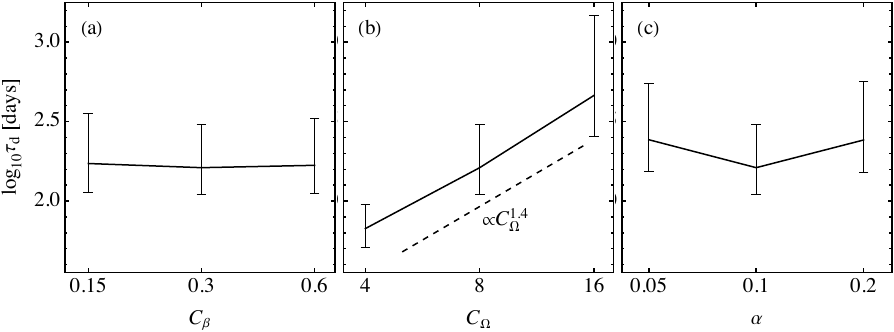}
\caption{Dependence of the damping time scale $\taud$ at $2500\text{\AA}$ on the dynamo parameters.}
\label{fig:scalings}
\end{figure*}

Based on the fiducial run with $\MBH=10^8\Msun$ and $\etaEdd=0.2$, we systematically vary one dynamo or disk parameter at a time to investigate how $\taud$ scales with each.
As shown in Figure~\ref{fig:scalings}, the results indicate that the magnetic field strength $\Cbeta$ and the viscosity parameter $\alpha$ have minor effects on $\taud$,
and the most significant dependency is with the dimensionless dynamo cycle period $\CO$ (with a slope of $\simeq1.4$),
which reflects the direct relationship between the dynamo time scale and accretion variability.
The relation can potentially be inverted to infer dynamo properties from AGN light curves.
Combined with Figure~\ref{fig:tau2500_MBH}, the results may suggest that the dynamo cycle periods can be systematically larger for lower-mass SMBHs, but it requires further simulations to confirm.

\subsection{Comparison with \cite{Lyubarskii1997}}
\label{sec:L97}

To conclude this section, we compare our model with the work by \cite{Lyubarskii1997} (hereafter L97).
The primary distinction between the L97 model and ours lies in the treatment of fluctuations:
L97 assumes they are spatially uncorrelated,
whereas we consider a physics-based model where the fluctuations are naturally correlated within a dynamo wavelength.
Specifically, in the L97 model, a stochastic component $\vartheta(t,r)$ is introduced into the viscosity parameter,
\beq
\alpha=\alpha_0[1+\vartheta(t,r)],
\eeq
where $\vartheta$ is a random variable with rms value $\vartheta_\text{rms}$, uncorrelated in $r$ but correlated in $t$ with a correlation time $\tau_\vartheta(r)$\footnote{%
In our simulations, the temporal correlation is implemented by generating a random number $\texttt{rand}\in[0,1)$ at each grid and each time step, and comparing it with $\texttt{dt}/\tau_\vartheta$, where \texttt{dt} is the current time step.
If the former is smaller, then the value of $\vartheta$ at location $r$ is updated.
In the case where \texttt{dt} is a constant, $\vartheta$ at a given $r$ follows a Poisson process.}.

The L97 model assumes that the value of $\tau_\vartheta$ is comparable to the viscous time $r^2/\nu=\alpha_0^{-1}\epsilon_0^{-2}\Omega_0^{-1}\tilde r^{5/4}$.
In our analysis, we parameterize it as $\tau_\vartheta(r)=\Clyu\Omega_0^{-1}\tilde r^{p_\text{L97}}$, allowing it to be shorter than the viscous time.
Furthermore, at each radius, $\vartheta$ is drawn from a uniform probability distribution in the range $[-0.02,0.02]$, yielding $\vartheta_\text{rms}\simeq0.012$.

\begin{figure*}
\centering
\includegraphics[width=0.75\textwidth]{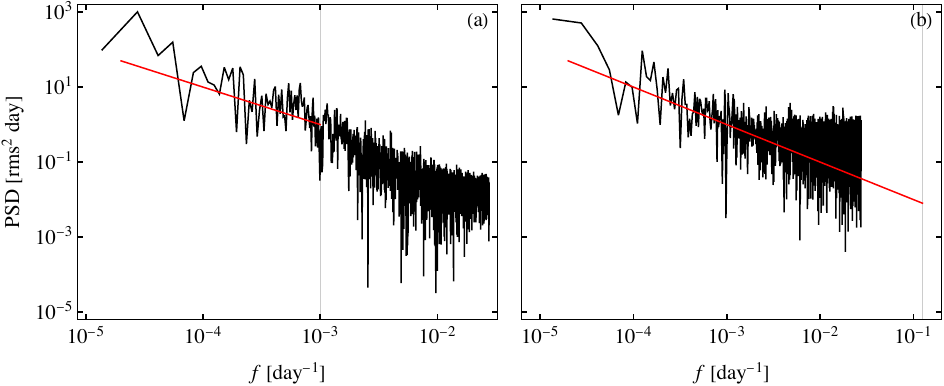}
\caption{PSDs of the accretion rates at the inner disk boundary
for runs~\texttt{L97V} (left) and \texttt{L97D} (right), based on the \cite{Lyubarskii1997} model.
The red lines indicate $\propto f^{-1}$,
and the vertical lines indicate the frequencies below which the PSDs are predicted to follow $\propto f^{-1}$ according to the L97 theory.}
\label{fig:l97_mdot_psd}
\end{figure*}

We consider two cases: (i) run~\texttt{L97V}, where $\Clyu=\alpha_0^{-1}\epsilon_0^{-2}$ and $p_\text{L97}=5/4$ corresponding to the original proposal of \cite{Lyubarskii1997} using the viscous time scale,
and (ii) run~\texttt{L97D}, where $\Clyu=8$ and $p_\text{L97}=3/2$, which yields a variability timescale roughly comparable to that of our fiducial dynamo model.
The PSDs of the accretion rates at the inner disk boundary are shown in Figure~\ref{fig:l97_mdot_psd}.
According to \cite{Lyubarskii1997}, the PSD of $\Mdot$ should scale as $f^{-1}$ at frequencies lower than $1/\tau_\vartheta(r_0)=1/C_\text{L97}$, reflecting the accumulation of multiplicative fluctuations from the outer disk.
We find that run~\texttt{L97V} agrees with the theoretical prediction well,
but run~\texttt{L97D} deviates from the $f^{-1}$ scaling in the middle to high frequency range.

\begin{figure*}
\centering
\includegraphics[width=0.444\textwidth]{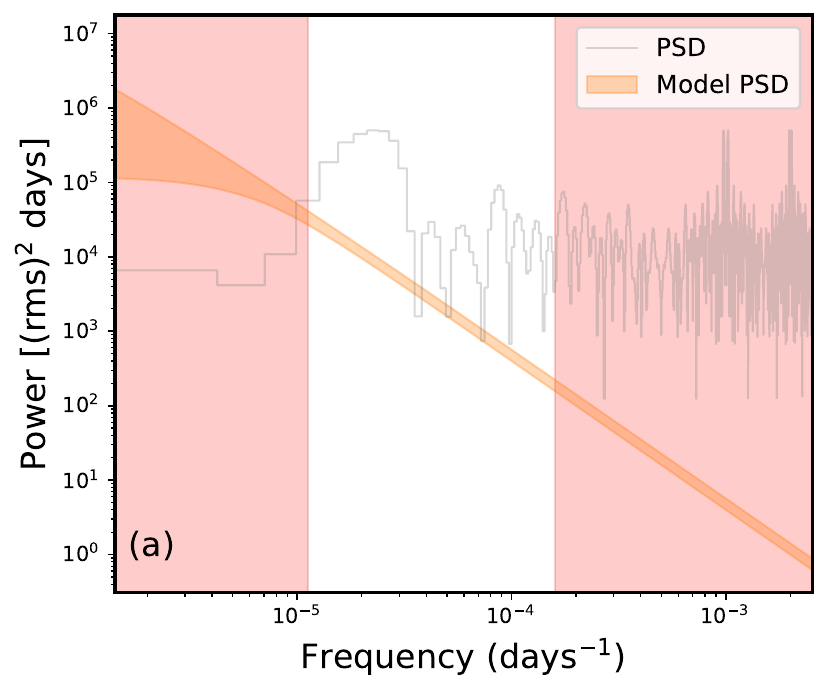}
\includegraphics[width=0.45\textwidth]{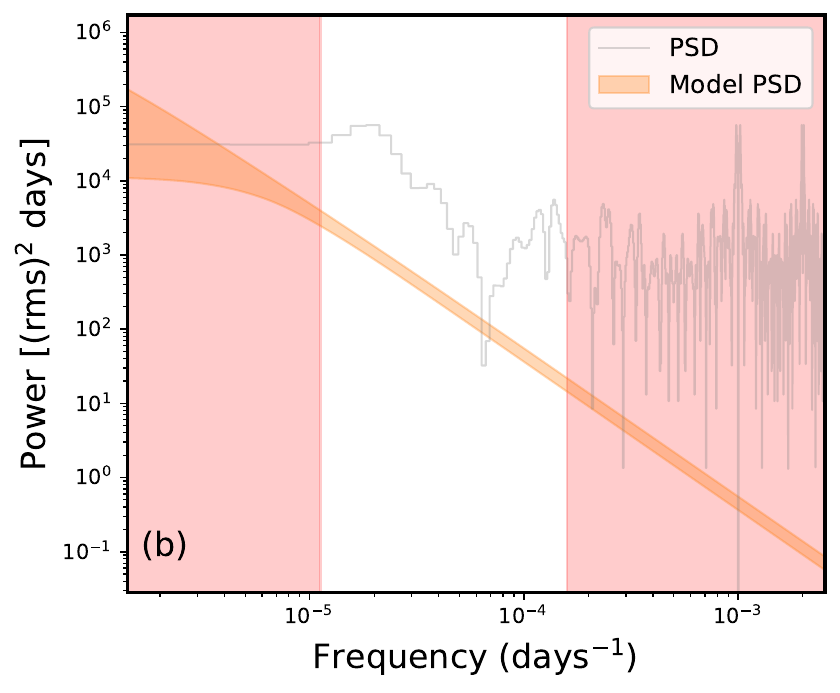}
\caption{Same as the bottom-right panel of Figure~\ref{fig:fid_tau_fit},
but for runs~\texttt{L97V} (left) and \texttt{L97D} (right) (see Figure~\ref{fig:l97_mdot_psd}).
The fitted damping times are both $\log_{10}\left(\taud/\text{days}\right)=4.8^{+0.6}_{-0.4}$.}
\label{fig:L97_2500A}
\end{figure*}

The PSDs and the fitted DRW models of the specific flux at $2500\text{\AA}$ are shown in Figure~\ref{fig:L97_2500A}.
For run~\texttt{L97V}, no clear low-frequency plateau is seen, or the simulation has not be run for long enough to display the plateau.
For run~\texttt{L97D}, the shape of the PSD marginally recovers that of a DRW model by eye, but the fitted damping time is $\gtrsim 10^{4.8}$ days, incompatible with observations.
The latter result suggests that simply assigning the correct variability time scale at each radius (by $C_\text{L97}$) is insufficient to reproduce the DRW time scale.
Compared to our dynamo-based model, we suggest that two additional factors may be crucial for  understanding the observations:
(i) the PSD of the fluctuation at each radius, where the L97 model assumes a white noise spectrum and our model exhibits a peaked spectrum, and
(ii) the spatial correlation of fluctuations, with the L97 model assuming no spatial correlation and our model incorporating a correlation length scale of $2\pi/\kD$.

\section{Conclusion \& Discussion}
\label{sec:conclusion}

Recent studies of AGN light curves have uncovered intriguing properties in their UV/optical variabilities.
In particular, the observed slowly moving disk fluctuations cannot be explained by a reverberation model.
In this work, we propose that quasi-periodic large-scale dynamos can be a possible source of driving fluctuations in disk accretion rates and emission profiles.
By modeling the disk dynamo as overlapping waves and exploring its consequences on accretion dynamics, we find that our model naturally yield slowly moving temperature patterns and a DRW-like PSD of the fluxes.
The scaling between the damping time and the wavelength transitions from $\propto \lambda$ to a plateau, and where the change occurs depends on the black-hole mass $\MBH$ and the Eddington ratio $\etaEdd$.
Furthermore, the scaling between the damping time at $2500\text{\AA}$ and $\MBH$ can reasonably recover the observation data at $\MBH\gtrsim 10^6\Msun$, but not below it, suggesting that additional factors may be considered such as the $\MBH$ dependence of dynamo properties.

Several works have reported on how the DRW damping time scales with AGN parameters, such as $\MBH$ and $\lambda$ \citep[e.g.,][]{MacLeod+2010,Burke+2021,Zhang+2024,Helias+2025}.
We provide a qualitative argument on why considerations beyond emission site geometry might be necessary for resolving this problem.
Consider two limiting cases regarding the emission site of photons at wavelength $\lambda$.
In the first case, we assume that they come from a narrow ring at radius $r_\lambda$ whose thermal emission peaks at $\lambda$ \citep[see also][]{MacLeod+2010}.
The surface temperature of the ring is $T\propto \lambda^{-1}$, and we have
\beq
\lambda \propto T^{-1}\propto D^{-1/4}\propto \MBH^{-1/4}\Mdot^{-1/4}r_\lambda^{3/4}.
\eeq
Hence
\beq
\taud
\propto \MBH^{-1/2}r_\lambda^{3/2}\propto \Mdot^{1/2}\lambda^2
\propto \eta^{-1/2}\eta_\text{Edd}^{1/2}\MBH^{1/2}\lambda^2,
\label{eqn:t_k}
\eeq
where $\eta=L/\Mdot c^2$.
In the other limit, we assume that photons at any wavelength have equal contributions from the whole disk, and thus $\taud$ loses its dependence on $\lambda$.
The only relevant time scale in the problem is then $\rg/c$, giving
\beq
\taud
\propto \rg/c\propto \MBH\lambda^0.
\eeq
In reality, we expect an intermediate scenario where photons at wavelength $\lambda$ originate from a finite region around $r_\lambda$.
Thus, we would expect a scaling exponent for the $\taud-\MBH$ relation between $1/2$ and $1$, and for $\taud-\lambda$ between $2$ and $0$.
Observationally, a value for the index of $\MBH$ of less than $1/2$ is commonly supported \citep{MacLeod+2010,Burke+2021}, which suggests that factors beyond a simplistic emission site geometry, such as magnetic fields, could influence the observed variability.

Large-scale magnetic fields in accretion disks can originate either from \textit{in situ} dynamo processes or through the advection of magnetic field lines from the surrounding environment.
For the former, Figure~\ref{fig:scalings} illustrates how the damping time scale is closely tied to the large-scale dynamo properties.
However, the precise dependence of dynamo parameters, such as cycle periods and coherence lengths, on both $\MBH$ and $\Mdot$ remains unclear and requires further exploration.
The different damping time scales in the low and high states of AGNs \citep{Ren+2024} may also be explained by the different operating dynamos in thin and thick disks.
Regarding advected fields, a greater advection efficiency would seemingly reduce the accretion variability by adding a non-fluctuating component to the disk field.
The efficiency of advection is influenced by several factors, including the coronal magnetic fields, the vertical structure of the disk, and disk winds \citep{BeckwithHawleyKrolik2009,GuiletOgilvie2012,GuiletOgilvie2013,CaoSpruit2013}.
Accounting for (i) how the dynamo processes are influenced by accretion physics,
and (ii) the role of advected fields, could help improve the agreement between our variability model and the observational data.

%Our model produces hard lags, i.e., the shorter-wavelength parts lag behind the longer-wavelength parts, similar to previous L97 models which utilizes fluctuation propagation \citep{ArevaloUttley2006}.
%\cite{Mushtukov+2018} has argued that fluctuation propagation can lead to both soft and hard lags when the Green's function is properly used, but the former is not observed in our model.
%We did not consider the hard X-ray reprocessing \cite{Krolik+1991,Nowak+1999,Nowak+1999b} and other outward-propagating models \citep[e.g.,][]{Cai+2018}.

\section*{Acknowledgment}

We thank Ying Zu, Eric Blackman and Mouyuan Sun for insightful discussions.
We thank the anonymous referee for detailed and constructive comments.
HZ acknowledges support from
the National Natural Science Foundation of China (No. 12403020),
Qimeng Project at Shanghai Polytechnic University,
and the China Postdoctoral Science Foundation (No. 2023M732251).
Part of the numerical simulations in this work were carried out on the Astro cluster supported by Tsung-Dao Lee Institute.

\bibliographystyle{aasjournalv7}
\bibliography{refs}

\end{document}